# Characteristic signatures of quantum criticality driven by geometrical frustration

Yoshifumi Tokiwa,[1,*,†] C. Stingl,[1,2] Moo-Sung Kim,[3] Toshiro Takabatake,[3] Philipp Gegenwart,[1,2]

[1] I. Physikalisches Institut, Georg-August-Universität Göttingen, 37077 Göttingen, Germany.
[2] Experimental Physics VI, Center for Electronic Correlations and Magnetism, University of Augsburg, 86159 Augsburg, Germany.
[3] Department of Quantum Matter, ADSM, Hiroshima University, Higashi-Hiroshima, 739-8530, Japan.

*Corresponding author; email: ytokiwa@scphys.kyoto-u.ac.jp
[†] Present address: Research Center for Low Temperature and Materials Science, Kyoto University, Kyoto 606-8501, Japan

## Abstract

Geometrical frustration describes situations where interactions are incompatible with the lattice geometry and stabilizes exotic phases such as spin liquids. Whether geometrical frustration of magnetic interactions in metals can induce unconventional quantum critical points is an active area of research. We focus on the hexagonal heavy fermion metal CeRhSn where the Kondo ions are located on distorted kagome planes stacked along the c axis. Low-temperature specific heat, thermal expansion and magnetic Grüneisen parameter measurements prove a zero-field quantum critical point. The linear thermal expansion, which measures the initial uniaxial pressure derivative of the entropy, displays a striking anisotropy. Critical and noncritical behaviors along and perpendicular to the kagome planes, respectively, prove that quantum criticality is driven by geometrical frustration. We also discovered a spin-flop-type metamagnetic crossover. This excludes an itinerant scenario and suggests that quantum criticality is related to local moments in a spin-liquid like state.

## MAIN TEXT

## Introduction

The ground state of 4f-based Kondo lattice systems depends sensitively on the balance of on-site Kondo screening of the moments and indirect exchange coupling between the latter, mediated by conduction electrons. Tuning this balance in controlled ways by changing composition, pressure or magnetic field allows to continuously change the ground state from long-range magnetically ordered to paramagnetic (*1*). Experimental studies have revealed a divergence of the quasi-particle mass and indications for a sudden reconstruction of the Fermi surface at the quantum critical point (QCP) possibly related to a localization of 4f electrons due to a destruction of Kondo screening (*2*). Theoretically, different scenarios for quantum criticality have been considered in a "global phase diagram" (*3-5*). In addition to the hybridization $J$ between 4f and conduction electrons, it considers a second important parameter $Q$, that is, the strength of quantum fluctuations. Large $Q$ is found for effective one-half spins on low-dimensional or highly frustrated lattices. In such situations, spin-liquid correlations among local 4f moments are competing with the Kondo singlet formation (*3-5*). Quantum criticality, arising by tuning $J$, is expected to be of conventional itinerant nature for small $Q$. Unconventional quantum criticality, due to a breakdown of the Kondo effect, is predicted for $Q$ being sufficiently large to suppress Kondo singlet formation when entering the magnetically ordered state. For very large $Q$, realized in highly geometrically frustrated materials, Kondo singlet formation breaks down even in the paramagnetic state. This may lead to a "fractionalized Fermi liquid" ground state which involves local fluctuating 4f moments that are effectively decoupled from the conduction electrons (*4*). The Fermi volume of this state does not include 4f degrees of freedom, in contrast to the case of ordinary heavy Fermi liquids. Such a metallic state with local 4f moments, which are not magnetically ordered but rather in a spin liquid ground state (*4,6*), has not yet been established experimentally. Furthermore, the proposal that strong geometrical frustration induces unconventional quantum criticality (*3,5*) needs to be exemplified.

A suitable candidate material is paramagnetic metal CeRhSn with hexagonal ZrNiAl structure. Its Ce-moments form a quasi-kagome lattice (*7*) with fully frustrated first-nearest neighbor interactions. Peculiar physical properties of isostructural heavy fermion materials such as YbAgGe (*8*) and CePdAl (*9,10*) indicate the importance of magnetic frustration in this structure type. The *T-H* phase diagram of YbAgGe is highly complex, displaying six different magnetic phases. CePdAl shows a partial magnetic order, where only two-thirds of the Ce-4f moments are ordered whereas one-third remain disordered (*9*). In contrast to these two systems, the single-ion Kondo scale of CeRhSn from bulk measurements is rather high, suggesting an intermediate valence nature of the 4f electrons (*7*). However, the highly anisotropic magnetic susceptibility which displays diverging behavior upon cooling to low temperatures is incompatible with the formation of Kondo singlets. Furthermore, the electrical resistivity exhibits non Fermi liquid (NFL) behavior (*7*). Strong antiferromagnetic fluctuations observed by $^{119}$Sn nuclear magnetic resonance (NMR) suggest that the system is close to a QCP (*11*). The absence of any magnetic order is confirmed down to 50 mK by μSR (muon spin rotation) experiments (*12*).

We consider the possibility that geometrical frustration within the quasi-kagome plane illustrated in Fig.1 is responsible for setting CeRhSn next to a QCP. In this case, highly anisotropic responses to uniaxial pressure along and perpendicular to the planes are expected. A deformation of the triangles by in-plane pressure along the a-axis, $p_a$, will lift the geometrical frustration. By contrast, pressure $p_c$ along the c-axis does not lead to such a deformation (because of the Poisson effect it only uniformly expands the in-plane triangular structure) and leaves geometrical frustration unchanged. The quantum critical contribution to the entropy should therefore be influenced dominantly by in-plane pressure. This can be detected by the linear thermal expansion, which equals the initial pressure derivative of the entropy along the measured direction.

A general hallmark of quantum criticality is divergent behavior in the Grüneisen ratio $\Gamma_r$=d (log(*T*)/d*r*)$_S$, which measures the temperature (*T*) contours at constant entropy (*S*) (*13,14*). Here *r* denotes the tuning parameter, for example, pressure or magnetic field. This generic divergence of $\Gamma_r$ results from the accumulation of entropy close to the QCP. At any pressure-sensitive QCP, the Grüneisen ratio $\Gamma=\beta/C$, given by the volume thermal expansion (β) divided by the specific heat (*C*), diverges (*13,14*). For the volume thermal expansion, the linear thermal expansion contributions along different main directions add up, for example, $\beta=2\alpha_a+\alpha_c$ for hexagonal CeRhSn. The ratio of the linear thermal expansion to the specific heat diverges only, if the quantum critical contribution of the entropy depends on the respective initial uniaxial pressure. As outlined above, for the quasi-kagome plane, this is not the case along the c-axis.

# Results

### Evidence of quantum criticality driven by geometrical frustration

Figure 2(A) displays the linear thermal expansion coefficient as $\alpha/T$, where $\alpha$ denotes the temperature derivative of the change of the sample length, measured in a high-resolution capacitive dilatometer (*15*). The coefficient along the c-axis saturates below 1K, whereas the in-plane one diverges upon cooling. Previous heat capacity measurements down to 0.5K have detected a saturation of the specific heat coefficient $C/T$ below 1K (*7*). Our data are in good agreement with the previous results. However, the data, measured down to milli-K temperatures, indicate that the system does not enter a Fermi liquid ground state. Rather, the heat capacity coefficient diverges upon cooling below 0.5K. NFL behavior has also been found in previous electrical resistivity (*7*) and nuclear magnetic resonance experiments (*11*). Next, we consider the Grüneisen ratio $\Gamma(T)$, determined using the volume thermal expansion and heat capacity. A clear divergence is found, signaling quantum criticality (*13*). Turning to the linear thermal expansion coefficient, we however, recognize a pronounced anisotropy. Although $\alpha_c/T$ is much larger than $\alpha_a/T$ above 0.2K, it is almost temperature independent, in contrast to the divergence found in $\alpha_a/T$. Anisotropic response of the linear thermal expansion for anisotropic materials is not unusual and has also been found, for example, in quantum critical tetragonal or orthorhombic heavy-fermion metals such as $CeNi_2Ge_2$ (*14*), $YbRh_2Si_2$ (*14*), $CeCu_{6-x}Ag_x$ (*16*) or $CeRhIn_{5-x}Sn_x$ (*17*). However, in all these cases, divergent behavior in $\alpha/T$ could be found along all main directions, indicating that any uniaxial pressure couples to quantum criticality. This is clearly different in CeRhSn (cf. Fig. 2A). Generally, the thermal expansion is given by the sum of normal (Fermi liquid) and quantum critical (NFL) contributions (*13*). Our data indicate a striking anisotropy of these two contributions: for CeRhSn the normal contribution is most pronounced along the c axis, whereas the quantum critical contribution is visible only along the a-axis. Note, that the absence of quantum criticality along the c axis could not be explained in terms of an ordinary single-ion anisotropy, because there is no proportionality between thermal expansion along the two different directions, that is, $\alpha_c$ should have been tiny compared to $\alpha_a$ then, which is clearly not the case. Consequently the data indicate that quantum criticality in CeRhSn only couples to uniaxial pressure along the a axis but not along the c direction. This unique observation is in agreement with the above consideration of a quantum critical state induced by geometrical frustration, where pressure along the c axis is not a relevant tuning parameter and leaves quantum criticality unchanged.

We now turn to the effect of an applied magnetic field. Field acts as relevant tuning parameter for Kondo lattice QCPs, because it forces the magnetic moments to align along the direction of the external field and thereby tunes the system away from quantum criticality. Indeed, a moderate field of 2 T leads to Fermi liquid behavior in the low-temperature specific heat (Fig. 2B). We probe the field-dependence of the quantum critical entropy accumulation by the measurement of the magnetic Grüneisen ratio $\Gamma_H = 1/T(\partial T/\partial H)|_S$. This property equals the adiabatic magnetocaloric effect and has been determined with the aid of an alternating-field technique (*18*). Similar as $\Gamma$ for any pressure sensitive QCP, $\Gamma_H$ diverges in the approach of any field-sensitive QCP (*13,19*). Indeed $\Gamma_H(T)$ diverges as $T \rightarrow 0$ at low fields in CeRhSn. Figure 3 shows data with the field applied along the c-axis, whereas similar behavior is also found for $H//a$. Because already a field of several tens of a mT leads to a crossover to Fermi liquid behavior [saturation of $\Gamma_H(T)$ for low $T$], the critical magnetic field must be very close to zero.

**Local moment metamagnetism**

Isothermal measurements of the magnetic Grüneisen ratio for the field along the a-axis are shown in Fig. 4(A). The divergence of $\Gamma_H$ toward zero field at $T$=0.1K again indicates a zero-field QCP. At elevated field near 3.6T, a further anomaly is observed: $\Gamma_H(H)$ shows a sharp sign change. A maximum found in specific heat at the same field confirms the bulk nature of this field-induced crossover (Fig. 4B). Upon raising the temperature this anomaly splits into two broad maxima, whose field dependence is indicated in the inset. Such behavior is indicative of metamagnetism, that is, the nonlinear sudden increase of magnetization in magnetic field. A metamagnetic crossover has, for example, been observed in $CeRu_2Si_2$ (20). Because the magnetic Grüneisen ratio can be expressed by $\Gamma_H = -(dM/dT)/C$, where $M$ is the magnetization, and the specific heat $C$ is always positive, the sign change of $\Gamma_H$ indicates a sign change of the temperature derivative of magnetization from $dM/dT > 0$ below 3.6T to $dM/dT < 0$ above. Such change is characteristic for metamagnetic crossovers, as found, for example, in heavy fermions and ruthenates (20-24).

Metamagnetism in anisotropic itinerant paramagnets such as $CeRu_2Si_2$ (20-22), $CeFe_2Ge_2$ (25), $UPt_3$ (26) and UCoAl (23) is observed to be most pronounced along the magnetic easy axis with the larger susceptibility, because it is related to the Zeeman splitting of a peak in the electronic density of states (27). The Landau expansion theory is used to explain this type of metamagnetism, which deals with the instability between paramagnetic and ferromagnetic states in a magnetic field (28,29). Rotating the field away from the easy toward the hard direction, which reduces the low-field susceptibility, therefore enhances the required field for metamagnetism.

Sharply distinct from itinerant metamagnets, CeRhSn displays a metamagnetic crossover only when the magnetic field is applied along the a axis, for which the susceptibility is ~30 times smaller than that along the c-axis (7). This excludes itinerant metamagnetism.

Another possibility is local moment metamagnetism which could be found in antiferromagnets. It is related to a spin flop along the Ising axis, that is, along the direction where the low-field magnetization is smallest (see the Supplementary Materials). In CeRhSn, as illustrated schematically in Fig. 5, strong geometrical frustration prevents long-range antiferromagnetic ordering. In the context of isostructural YbAgGe, which displays quantum-bicritical behavior, we have previously discussed the evolution with increasing frustration $Q$ from a first-order metamagnetic transition through such a quantum bicritical point toward a metamagnetic crossover within a spin liquid regime (8). In CeRhSn the combination of zero-field quantum criticality and a metamagnetic crossover along the a axis is consistent with the suppression of local moment magnetic order by frustration (cf. Fig. 5). Such a spin liquid state of local f electrons in metallic environment has been discussed as fractionalized Fermi liquid in the global phase diagram for frustrated Kondo lattices (3-5).

# Discussion

We have found thermodynamic evidence for a zero-field QCP by performing a Grüneisen ratio analysis. Furthermore, the observed metamagnetic crossover along the axis with smaller low-field magnetization could not be described by an itinerant scenario but rather points to the presence of local moments. However, these moments are not ordered at zero temperature, but in a quantum critical state. In combination with the observed anisotropic behavior of the thermal expansion these observations indicate that quantum criticality is related to the strongly frustrated arrangement of 4f moments in CeRhSn. Several remarks are in order. First, our observation of quantum critical behavior in thermal expansion only along one direction is distinctly different from the previous studies of the Grüneisen ratio divergence near a QCP, where thermal expansion is singular along all directions (*14,16,17*). In general, anisotropic thermal expansion may also be related to single ion anisotropy effects which could result in very small expansivity along certain directions. In our case $\alpha_c$, which lacks signatures of quantum criticality, is however large compared to $\alpha_a$, implying that there is no such trivial explanation for the absence of quantum criticality in $\alpha_c$. Second, previous observation of NFL behavior in the low-*T* electrical resistivity and magnetic susceptibility of CeRhSn has been discussed in relation to possible atomic disorder (*7,30*). As detailed in Materials and Methods, the characterization of the investigated single crystal did not reveal indication of such disorder. Furthermore, $^{119}$Sn NMR showed no distribution of relaxation and no satellites, excluding a large amount of disorder, and led to a conclusion that a disorder driven mechanism fails to explain the NFL behavior (*11*). We also note that at most logarithmic divergence of the Grüneisen ratios is allowed for a smeared quantum phase transition or quantum Griffiths singularity (*14,31*). The observed power-law divergence of the thermal and magnetic Grüneisen ratios unambiguously proves that quantum criticality is realized, that is, that the correlation length and time of magnetic order parameter fluctuations diverge in the approach of zero temperature and zero magnetic field. Third, if quantum criticality would be related to a possible valence change, resulting in a quantum critical end point (*32*), one would have expected strongly divergent behavior in $\alpha/T$ along all directions. Such scenario would also be incompatible with the observed metamagnetism. Fourth, entropy involving quantum critical behavior below 0.5 K is estimated by integrating *C/T* and found to be tiny, ~0.02*R*ln(2). This indicates a reduced size of fluctuating moments in CeRhSn. Such strong reduction is probably caused both by Kondo screening and quantum fluctuations due to geometrical frustration. The former effect is found in heavy-fermion materials, such as YbRh$_2$Si$_2$ (33). The latter is found, for example, in the highly frustrated magnet Li$_2$IrO$_3$ (34). Fifth, the observed insensitivity of quantum criticality to initial c-axis uniaxial pressure, which retains the geometrical frustration, suggests a quantum critical phase in CeRhSn. Measurements under true uniaxial pressure of the order of kilobars, which is sufficient to substantially modify geometrical frustration and presumably induce long-range magnetic order would be very interesting. Sixth, quantum criticality without fine-tuning of composition or pressure has also been found in Pr$_2$Ir$_2$O$_7$ (*35*) and β-YbAlB$_4$ (*36*), which both have frustrated magnetic interactions, and an extended gapless spin liquid phase has recently been proposed for EtMe$_3$Sb[Pd(dmit)$_2$]$_2$ (*37*) but for neither of these materials data of the low-*T* linear thermal expansion are yet available.

To summarize, the paramagnetic quasi-kagome Kondo lattice CeRhSn exhibits intriguing properties, which we ascribe to quantum criticality induced by geometrical frustration. Divergent Grüneisen ratios Γ and Γ$_H$ provide evidence for a zero-field QCP. The anisotropy of thermal expansion, displaying quantum critical behavior only along the a axis but not along the c axis, indicates that geometrical frustration drives quantum criticality in this system. Despite the large Kondo scale of $T_K$~200K (*7*), the presence of local magnetic moments is suggested by spin flop metamagnetism at low temperatures. From the perspective of the proposed global phase diagram for Kondo lattice systems (*3-5*), CeRhSn appears as a candidate for the realization of a metallic quantum spin liquid in a frustrated Kondo lattice.

# Materials and Methods

### Study design

This study was designed for discovering characteristic signatures of a QCP driven by geometrical frustration. To examine the presence of quantum criticality and the characteristic signatures of such a QCP in the frustrated Kondo lattice CeRhSn, growth of single crystalline samples, careful examination of the sample quality, specific heat, thermal expansion and magnetocaloric effect measurements at low temperatures down to ~70mK were carried out.

**Sample preparation and examination of sample quality**

An ingot of a CeRhSn single crystal was grown by the Czochralski method from the melt of stoichiometric amounts of the constitute elements in a radio frequency induction furnace. The crystal was wrapped with tantalum foil, sealed in an evacuated quartz tube, and annealed at 900°C for 3 weeks.

For materials very close to a QCP, physical properties at low temperatures can be extremely sensitive to crystalline disorder, which also can induce local distortions. Indeed for polycrystalline samples of CeRhSn, spin glass behavior was found and annealing of the samples changed the physical properties markedly (*7*). Furthermore, a clear specific heat anomaly was observed at 6.2K. In our single crystalline samples on the other hand, neither spin glass behavior nor an anomaly at 6.2 K is present (*27*), indicating a much better quality. It also proves that the transition at 6.2 K is either caused by a foreign phase or induced by structural disorder. Moreover, we do not observe any appreciable difference in the physical properties of as-grown and annealed samples (*38*). This suggests that our crystals are well structurally ordered already without annealing. It has also been reported that the system is very sensitive to small changes in chemical composition (*7*). Within the 1% accuracy of electron microprobe analysis our crystals are homogeneous and there is no deviation from the nominal stoichiometry. We used single crystalline samples of the same ingot, as in the previous study (*7*). Specific heat of two different pieces was measured and confirmed no sample dependence. For thermal expansion a larger piece of the ingot (~0.6g) has been investigated.

**Low-temperature experiments**

The linear thermal expansion was measured using a high-resolution capacitive dilatometer (*15*) adapted to a dilution refrigerator. Here, the sample is fixed between a movable part and outer frame by two parallel flat springs that exert a small force on the sample along the measurement direction. The amount of force applied in the particular dilatometer and in the capacitance range of our measurement is 2.6 N. Two parallel surfaces of the sample were pressurized. The surface areas of measurements along the c axis and a axis are $1.3 \times 10^{-5} m^2$ and $1.2 \times 10^{-5} m^2$, respectively. The uniaxial pressure applied on the sample is then 2.0 and 2.2 bars for c axis and a axis, respectively. The two surfaces are perfectly parallel to each other and the force is applied perpendicular to the surfaces, so that there is no differential contraction and no sheer strain on the sample. All detected length changes are fully reversible and reproducible without any indication for sample cracks induced by strain or any other irreversibilities.

The magnetocaloric effect and specific heat were measured using an alternating field technique (*16*) and the standard quasi-adiabatic heat pulse method, respectively, adapted to a dilution refrigerator.

**Acknowledgments**

We acknowledge collaboration with M. Brando and M. Garst, as well as, stimulating discussions with P. Coleman, M. Garst, Y. Matsuda and Q. Si.
**Funding:** This work has been supported by the German Science Foundation through FOR 960 (Quantum phase transitions) and the Helmholtz Virtual Institute 521.
**Author contributions:** Y. T. conceived the project. Y. T. and P. G. planned and designed the experiments. M.S.K. and T.T. synthesized and characterized the samples. Y.T. and C.S. performed the experiments. Y.T. and P.G. discussed the results and prepared the manuscript.
**Competing interests:** The authors declare that they have no competing interests.

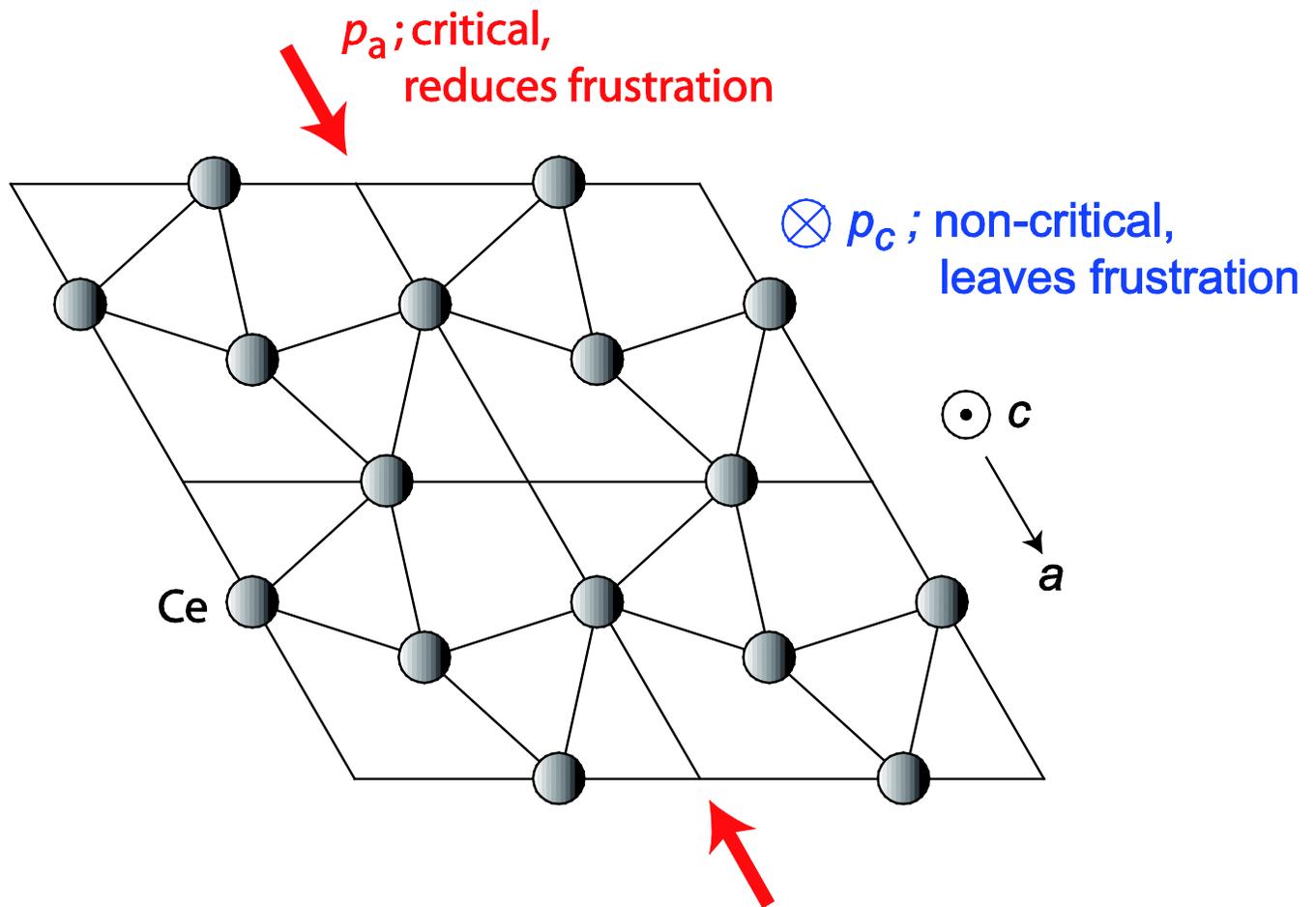

**Fig. 1. Critical and noncritical control parameters for a two dimensional geometrically frustrated lattice of magnetic moments.** Black circles indicate Ce atoms in the quasi-kagome plane of CeRhSn. Uniaxial pressure parallel to the a axis ($p_a$) deforms the equilateral triangular units and will reduce the geometrical frustration, whereas pressure perpendicular to the plane ($p_c$) leaves frustration unchanged.

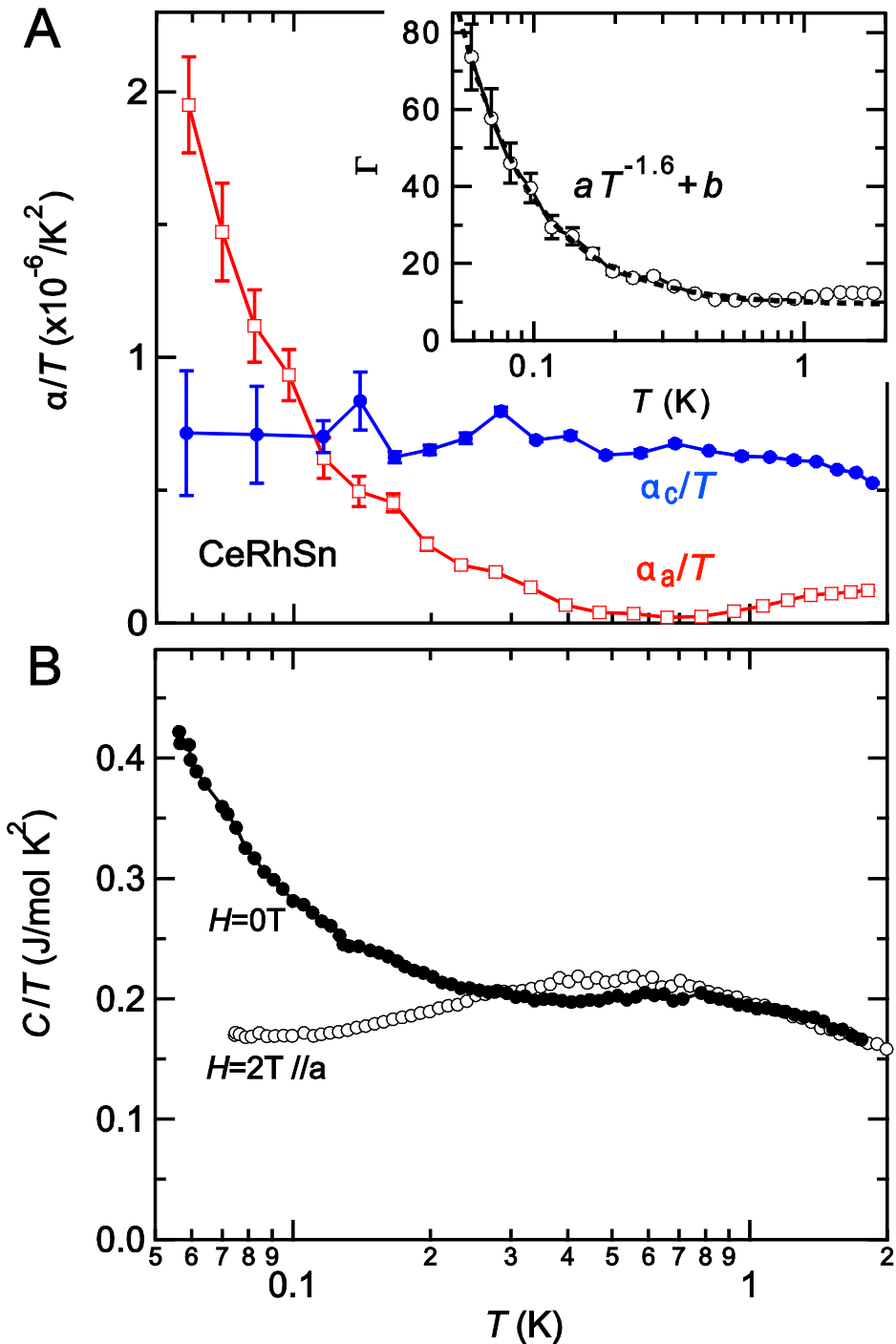

**Fig. 2. Evidence of a frustration-induced QCP in CeRhSn.** (**A**) Thermal expansion coefficient divided by temperature $\alpha/T$ versus temperature measured along the a- and c directions, as indicated by red and blue symbols. The inset shows the Grüneisen parameter $\Gamma = BV_m\beta/C$, where $B=105$ GPa is the bulk modulus of isostructural UCoAl (*39*), $V_m = 1.36 \times 10^{-4}$ m$^3$/mol is the molar volume, $\beta = 2\alpha_a + \alpha_c$ is the volume thermal expansion and $C$ denotes specific heat. The black dotted line indicates a power-law divergence. (**B**) Specific heat divided by temperature $C/T$ as a function of temperature. Solid and open circles indicate data measured at zero field and at 2 T applied parallel to the a-axis, respectively. Up to 2 T there is no appreciable nuclear Schottky contribution visible even at lowest measured temperatures.

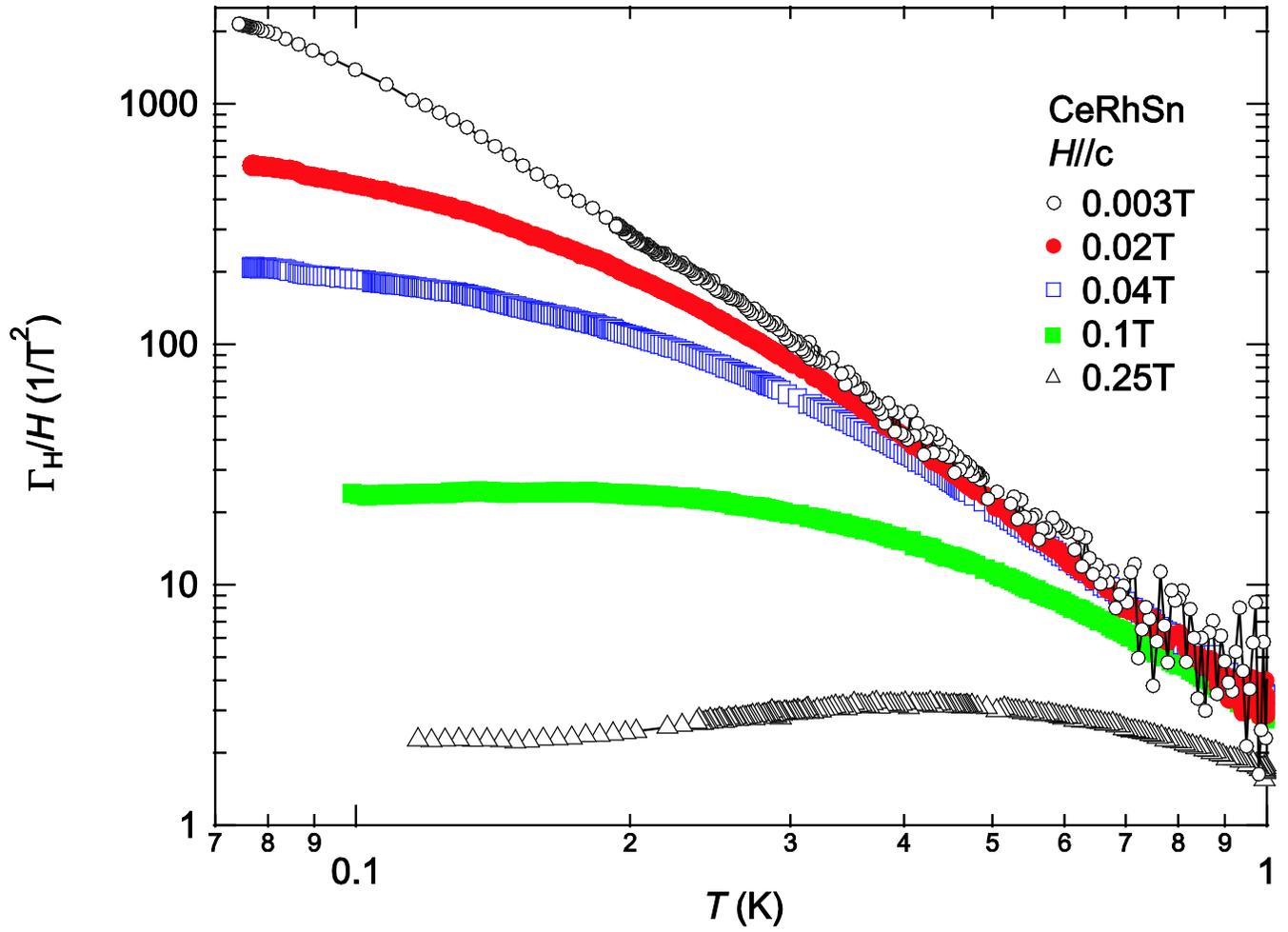

**Fig. 3. Divergence of the magnetic Grüneisen parameter $\Gamma_H$.** $\Gamma_H/H$ as a function of temperature at various magnetic fields applied parallel to the c axis.

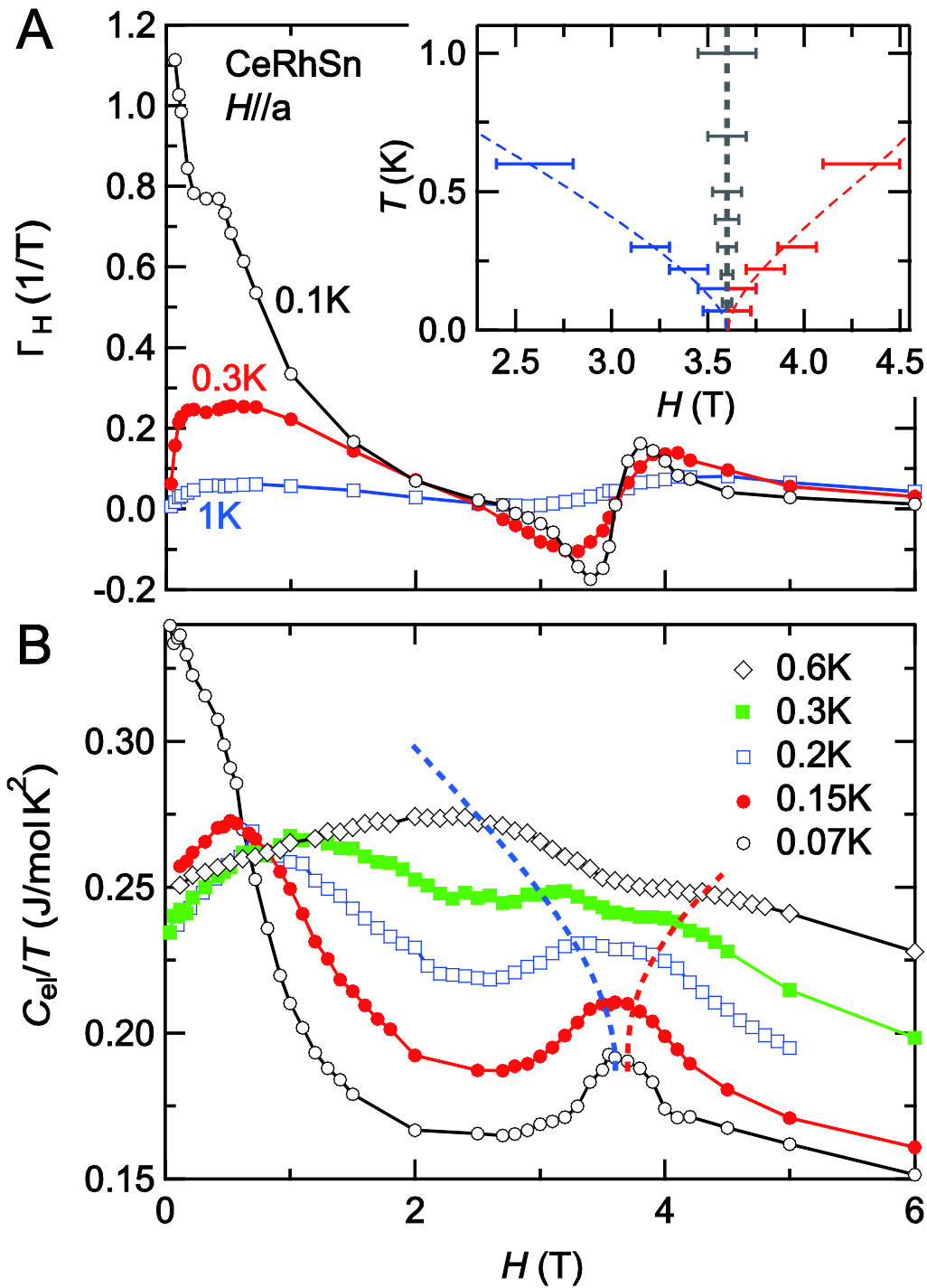

**Fig. 4. Spin flop crossover in the spin liquid state of CeRhSn.** (**A**) Magnetic Grüneisen ratio $\Gamma_H$ as a function of magnetic field applied parallel to the a axis. The inset shows a temperature versus magnetic field phase diagram, where the gray dotted line has been obtained from inflection points of $\Gamma_H(H)$ and the blue and red lines indicate anomalies in the field dependence of the electronic specific heat. (**B**) Electronic specific heat divided by temperature $C_{el}/T$ versus magnetic field applied parallel to the a-axis at constant temperatures. The nuclear contribution has been subtracted from the raw data. $C_{el}/T$ data at 0.15, 0.2, 0.3 and 0.6 K are shifted vertically for clarity by 0.15, 0.3, 0.45 and 0.6 J/mol K$^2$, respectively. The blue and red dotted lines indicate the positions of specific heat anomalies.

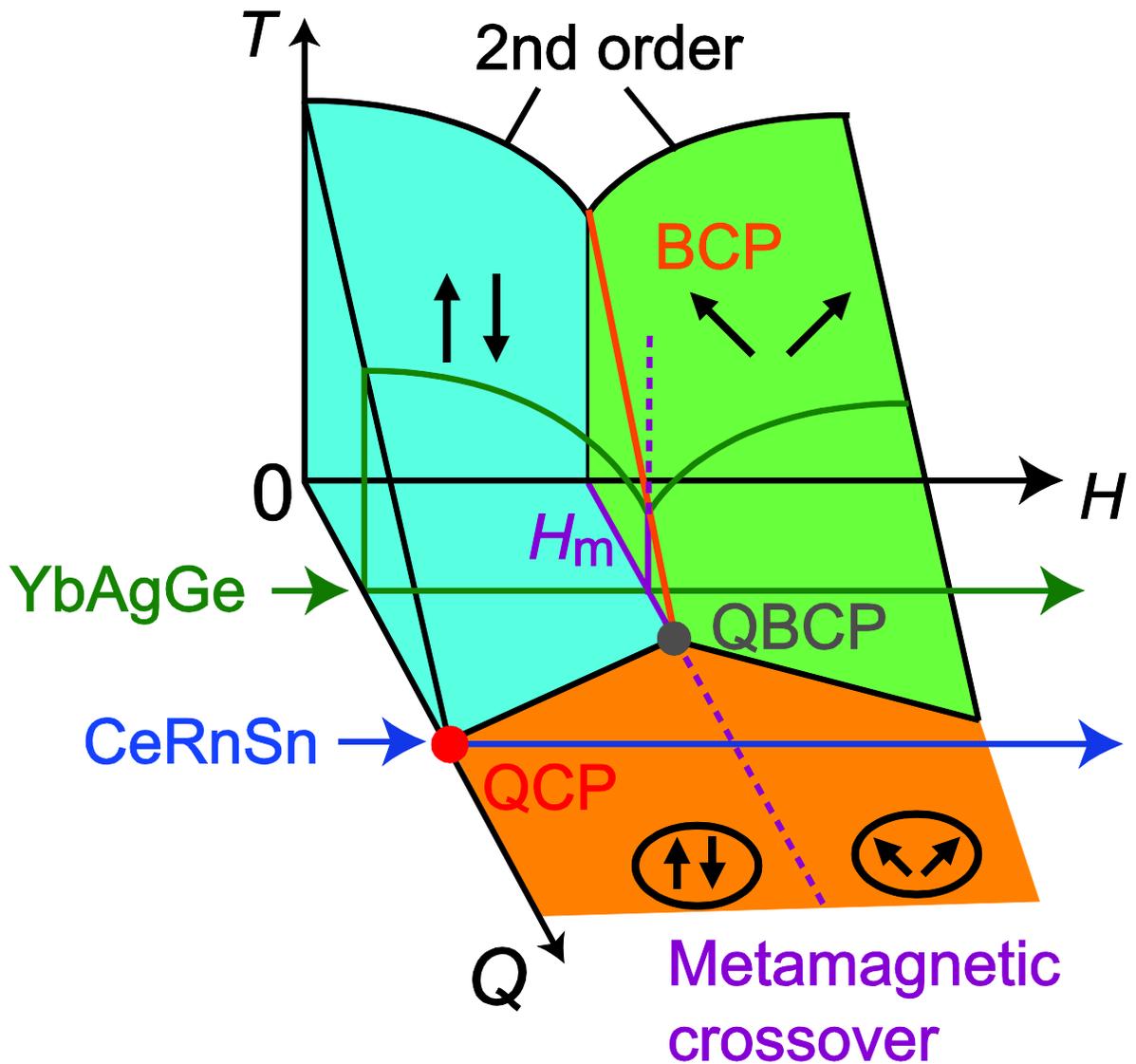

**Fig. 5. Possible scenario for the metamagnetic crossover in CeRhSn.** $T$-$H$-$Q$ phase diagram with a spin-flop transition between two magnetically ordered states (*8*). The parameter $Q$ indicates the strength of quantum fluctuations induced by geometrical frustration. A line of bicritical points (BCP, in red) separates two distinct magnetically ordered states. The quantum bicritical point (QBCP) is the point where $T_{BCP}$ approaches zero temperature. The purple solid and dotted lines indicate $H_m$ and the metamagnetic crossover, respectively. The isostructural heavy fermion antiferromagnet YbAgGe is positioned near the QBCP and its $T$-$H$ phase diagram near $H_m$ is illustrated by solid green lines (*8*). In this material, field induced quantum critical behavior arises from the nearby QBCP. Paramagnetic CeRhSn displays a zero-field QCP induced by $Q$ (solid red circle) and field-driven metamagnetic crossover (dotted purple line) of spin flop nature.

# Supplementary Materials

**Local-moment and itinerant metamagnetism**

The metamagnetic magnetization processes for local ordered moments and paramagnetic itinerant moments are illustrated in Figs. S1(A) and (B), respectively. For local moments, the magnetization along the Ising axis is zero up to a critical field, $H_m$, where it increases abruptly. Above $H_m$, the Zeeman energy overcomes the Ising anisotropy energy. When the magnetic field is applied perpendicular to the Ising axis, the moments are continuously canted towards the field direction and the magnetization increases continuously up to saturation. On the other hand, itinerant metamagnetism for materials close to the border of ferromagnetism is observed preferentially along the direction of larger magnetization. Tuning the field out of this direction shifts $H_m$ to higher fields (26). If metamagnetism, like in CeRhSn, is only found along the hard- but not along the easy axis, it could therefore not be of itinerant nature.

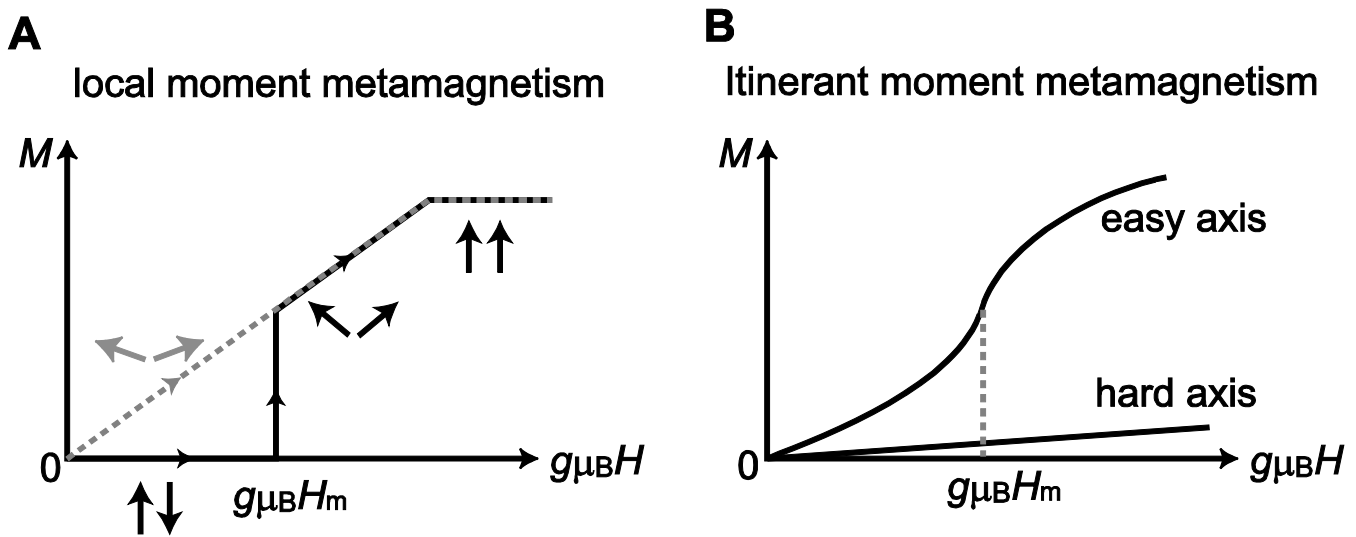

**Fig.S1**. **Illustration of local-moment and itinerant metamagnetism.** (A) Magnetization process of local ordered moments with Ising anisotropy. Black solid and grey dotted lines indicate the magnetization along and perpendicular to the Ising axis, respectively. (B) Respective magnetization behavior for itinerant metamagnetism. Note, that metamagnetism is found along the easy direction of larger initial susceptibility, while along the hard axis much larger fields are requires to achieve a respective magnetic polarization.